\documentclass[aps,prd,preprint,superscriptaddress,amsmath,amssymb,showpacs]{revtex4-1}
\usepackage{todonotes}
\usepackage{dcolumn}
\usepackage{graphicx}
\usepackage{float}
\usepackage{physics}
\usepackage[colorlinks=true,allcolors=blue]{hyperref}
\usepackage{epstopdf}
\usepackage{changes}
\usepackage{xcolor}

\begin{document}

\title{Neural Network Modeling of Heavy-Quark Potential from Holography}

\author{Ou-Yang Luo}
\affiliation{School of Nuclear Science and Technology University of South China Hengyang, China No 28, West Changsheng Road, Hengyang City, Hunan Province, China.}

\author{Xun Chen}
\email{chenxunhep@qq.com}
\affiliation{School of Nuclear Science and Technology University of South China Hengyang, China No 28, West Changsheng Road, Hengyang City, Hunan Province, China.}
\affiliation{Key Laboratory of Quark and Lepton Physics (MOE), Central China Normal University, Wuhan 430079,China.}

\author{Fu-Peng Li}
\email{fupengli29@mails.ccnu.edu.cn}
\affiliation{Key Laboratory of Quark and Lepton Physics (MOE) and Institute of Particle Physics, Central China Normal University, Wuhan 430079, China.}

\author{Xiao-Hua Li}
\email{lixiaohuaphysics@126.com}
\affiliation{School of Nuclear Science and Technology University of South China Hengyang, China No 28, West Changsheng Road, Hengyang City, Hunan Province, China.}

\author{Kai Zhou}
\email{zhoukai@cuhk.edu.cn}
\affiliation{School of Science and Engineering, The Chinese University of Hong Kong, Shenzhen (CUHK-Shenzhen), Guangdong, 518172, China.}

\begin{abstract}
Using Multi-Layer Perceptrons (MLP) and Kolmogorov-Arnold Networks (KAN), we construct a holographic model based on lattice QCD data for the heavy-quark potential in the 2+1 system. The deformation factor $w(r)$ in the metric is obtained using the two types of neural network. First, we numerically obtain $w(r)$ using MLP, accurately reproducing the QCD results of the lattice, and calculate the heavy quark potential at finite temperature and the chemical potential. Subsequently, we employ KAN within the Andreev-Zakharov model for validation purpose, which can analytically reconstruct $w(r)$, matching the Andreev-Zakharov model exactly and confirming the validity of MLP. Finally, we construct an analytical holographic model using KAN and study the heavy-quark potential at finite temperature and chemical potential using the KAN-based holographic model. This work demonstrates the potential of KAN to derive analytical expressions for high-energy physics applications.
\end{abstract}

\maketitle
\section{INTRODUCTION}
\label{sec-int}
Quantum Chromodynamics (QCD) serves as the fundamental theory describing the interactions of quarks and gluons, the basic constituents of matter in the strong force realm. Despite its foundational role, an accurate description of real-world QCD, especially under conditions that deviate from the idealized scenarios often considered in theoretical models, remains an elusive goal for physicists. The AdS/CFT correspondence \cite{Maldacena:1997re}, also known as the Anti-de Sitter/Conformal Field Theory duality, is a powerful theoretical tool that has significantly impacted the study of QCD. At present, numerous "top-down" methodologies are employed to extract realistic representations of holographic QCD from string theory \cite{Burrington:2004id,Sakai:2005yt,Sakai:2004cn,BitaghsirFadafan:2018uzs,BitaghsirFadafan:2019ofb,Abt:2019tas,Nakas:2020hyo,
Fujita:2022jus,Yadav:2023glu,Li:2015uea,Li:2015kma,Li:2024apc,Bigazzi:2024sjy}. On the other hand, "bottom-up" approaches focus on evaluating holographic QCD models by utilizing experimental data and results from lattice calculations \cite{Andreev:2006nw,He:2007juu,Braga:2017fsb,Ferreira:2019nkz,Chen:2019rez,Arefeva:2024xmg,Chen:2021bkc,Guo:2023zjx,Fang:2015ytf,Chen:2018vty,
Chen:2020ath,Li:2024lrh,Fu:2024wkn,Bea:2024xgv,Jokela:2024xgz,Wang:2024rim,Wang:2024szr,Zhao:2022uxc,Chen:2022obe,Caldeira:2021izy,Arefeva:2018hyo,
Zhu:2021nbl,Wen:2024hgu,Liang:2023lgs,Cao:2022csq,Cao:2022mep,McInnes:2018ibt,Sonnenschein:2024rzw,DeWolfe:2010he,Yang:2014bqa,Li:2012ay,Chen:2021gop,
Brodsky:2014yha,Li:2021cwv}.

In the experimental analysis of quark-gluon plasma and its characteristics, heavy quarks are examined with heightened sensitivity. These heavy quarks act as crucial probes for identifying the presence of QCD matter at finite temperatures \cite{Matsui:1986dk,Chen:2024iil,Yang:2015aia,Zhou:2020ssi, Zhou:2014kka}. The dissociation of heavy quark-antiquark pairs is commonly acknowledged as an indication of deconfinement-induced color screening, which makes the quark-antiquark potential a subject of significant interest within the realm of holographic QCD. The holographic potential for quark-antiquark pairs was originally reported in \cite{Maldacena:1998im}. Investigating how quarks are held together within hadrons allows researchers to glean insights into the strong interaction forces, thereby unveiling the complexities of the subatomic world governed by QCD. The heavy-quark potential has been investigated in various holographic QCD models in recent years \cite{Andreev:2006ct,Andreev:2006eh,He:2010bx,Colangelo:2010pe,Li:2011hp,Fadafan:2011gm,Fadafan:2012qy,Cai:2012xh,Zhang:2015faa,
Ewerz:2016zsx,Chen:2017lsf,Bohra:2019ebj,Zhou:2023qtr,Giataganas:2011nz}.

Multi-layer perceptrons (MLP) have made numerous remarkable achievements in addressing inverse and variational problems in various scientific domains, credited to their robust representational capability \cite{gupta2022aitheoreticalparticlephysics,Zhou:2023pti,He:2023zin,boehnlein2022colloquium,thuerey2021physics,Wang:2021jou,Shi:2022yqw}. This capability is underpinned by the universal approximation theorem \cite{hornik1989multilayer}, which posits that multi-layer feed-forward neural networks, when equipped with a sufficient tally of hidden neurons, can approximate any well-behaved functions. One notable implementation of MLP is their use in representing solutions to partial differential equations (PDEs) \cite{raissi2019physics,Soma:2022vbb,karniadakis2021physics,Shi:2021qri}. Recently, Kolmogorov-Arnold Networks (KAN) have been proposed in Ref. \cite{Liu:2024swq}, and shown to outperform MLP in terms of interpretability for small-scale natural science AI tasks.

The application of deep learning (DL) to holographic QCD has been explored in recent years, since the seminal work \cite{Hashimoto:2018ftp}. Furthermore, the integration of machine learning with holographic QCD has been extensively explored in a range of recent studies, as evidenced by the contributions of \cite{Akutagawa:2020yeo, Hashimoto:2018bnb, Yan:2020wcd, Hashimoto:2021ihd, Song:2020agw, Chang:2024ksq, Ahn:2024gjf, Gu:2024lrz, Li:2022zjc, Cai:2024eqa, Ahn:2024lkh, Mansouri:2024uwc, Chen:2024mmd, Jejjala:2023zxw}. Unlike conventional holographic models, this approach first employs experimental or lattice QCD data to determine the bulk metric and other model parameters through machine learning. Subsequently, the determined metric is utilized to calculate other physical QCD observables, which delivers predictions of the model.

The rest of our paper is organized as follows. In Sec. \ref{sec2}, the general holographic method of calculating the heavy-quark potential is introduced. The MLP are used to numerically derive the deformed factor $w(r)$ and calculate the potential at finite temperature and chemical potential in Sec. \ref{sec3}. In Sec. \ref{sec4}, KAN is used to confirm the validity by reproducing the Andreev-Zakharov model and is used to analytically derive $w(r)$ from lattice results. We compare the MPLs, KAN, and Andreev-Zakharov model in Sec. \ref{sec5}. Finally, we give a summary in Sec. \ref{sec6}.

\section{Holographic heavy-quark potential}\label{sec2}
In the original paper by Maldacena \cite{Maldacena:1998im}, he derived the Coulombic potential for the heavy-quark potential within the framework of \(\mathcal{N}=4\) Super Yang-Mills theory. Later, Andreev and Zakharov introduced a deformation factor to break the conformal symmetry, thereby reproducing the correct behavior of the heavy-quark potential in the holographic model \cite{Andreev:2006nw}. Even though the Andreev-Zakharov model is a phenomenological model, it can capture the behavior of the heavy-quark potential. Recently, this model has been extended to calculate the potential of exotic states, as shown in a series of works \cite{Andreev:2012mc, Andreev:2015riv, Andreev:2019cbc, Andreev:2020xor, Andreev:2021eyj, Andreev:2022qdu, Andreev:2023hmh, Andreev:2024orz}. The background metric can be expressed as
\begin{equation}
\begin{aligned}
ds^2 &= w(r)\frac{1}{r^2} \bigl[- f(r) dt^2 + d\vec{x}^2 + f^{-1}(r) dr^2 \bigr], \\
f(r) &= 1 - \left(\frac{1}{r_h^4} + q^2 r_h^2\right) r^4 + q^2 r^6.
\end{aligned}
\end{equation}
$q$ is the black hole charge, $r_h$ is the position of the black hole horizon. The Hawking temperature of the black hole is defined as
\begin{equation}
    T = \frac{1}{4\pi} \left| \frac{df}{dr} \right|_{r=r_h} = \frac{1}{\pi r_h} \left(1 - \frac{1}{2}Q^2\right),
\end{equation}
where \(Q = qr_h^3\) and \(0 \leq Q \leq \sqrt{2}\). The relationship between the chemical potential \(\mu\) and \(q\) is given as
\begin{equation}
    \mu = k \frac{Q}{r_h}.
\end{equation}
\(k = 1\) is a dimensionless parameter, and we fix the parameter \(k\) to one in this paper.
Thus, we can get
\begin{equation}
    \label{eq:frt}
    f(r) = 1 - \left(\frac{1}{r_h^4} + \frac{\mu^2}{r_h^2}\right) r^4 + \frac{\mu^2}{r_h^4} r^6,
\end{equation}
\begin{equation}
    T = \frac{1}{\pi r_h} \left(1 - \frac{1}{2} \mu^2r_h^2 \right).
\end{equation}
If we choose the static gauge \(\tau = t\), \(\sigma = x\), then a static quark-antiquark pair locating at
\begin{equation}
    x(z=0) = -\frac{L}{2},\quad \text{and} \quad x(z=0) = \frac{L}{2}.
\end{equation}
The Nambu-Goto action of the U-shaped string can be expressed as
\begin{equation}
  S = \frac{1}{2\pi\alpha'} \int d\tau d\sigma \sqrt{-\det(g_{\alpha\beta})},
\end{equation}
with
\begin{equation}
g_{\alpha\beta} = G_{\mu\nu} \frac{\partial x^{\mu}}{\partial \sigma^{\alpha}} \frac{\partial x^{\nu}}{\partial \sigma^{\beta}}.
\end{equation}
The action can now be written as
\begin{equation}
  S = \frac{g}{T} \int_{-\frac{L}{2}}^{\frac{L}{2}} dx \, {\frac{w(r)}{r^2}} \sqrt{f(r) + (\partial_x r)^2},
\end{equation}
where $g=\frac{1}{2\pi\alpha'}$ is related to the string tension.
Now we identify the Lagrangian as
\begin{equation}
  \mathcal{L} = \frac{w(r)}{r^2}  \sqrt{f(r) + (\partial_x r)^2}.
\end{equation}
Then we have
\begin{equation}
  \mathcal{H} = \frac{w(r)f(r)}{r^2\sqrt{f(r) + (\partial_x r)^2}}.
\end{equation}
At points $r_0$, we have
\begin{equation}
  \frac{w(r)f(r)}{r^2\sqrt{f(r) + (\partial_x r)^2}} = \frac{w(r_0)}{r_0^2}\sqrt{f(r_0)}.
\end{equation}
As before, $\partial x_r$ can be solved as
\begin{equation}
  \partial_x r = \sqrt{\frac{w^2(r)f^2(r)/r^4 - w^2(r_0)f(r_0)f(r)/r_0^4}{w^2(r_0)f(r_0)/r_0^4}}.
\end{equation}
Thus, $\partial r_x$ is
\begin{equation}
  \partial_r x = \sqrt{\frac{w^2(r_0)f^2(r_0)/r_0^4}{w^2(r)f^2(r)f(r_0)/r^4 - w^2(r_0)f^2(r_0)f(r)/r_0^4}}.
\end{equation}
The distance \( L \) of the quarks is given by
\begin{equation}
    \label{L}
    L = 2 \int_0^{r_0} \partial_r x \, dr =  2 \int_0^{r_0} \sqrt{\frac{f(r_0)w^2(r_0)/r_0^4 }{f^2(r) w^2(r)/r^4  - f(r_0) f(r) w^2(r_0)/r_0^4 }} dr.
  \end{equation}
Subtracting the divergent term at $z = 0$, the heavy-quark energy can be written as
\begin{equation}
  \label{E_eq}
  \begin{aligned}
    E &= 2g \int_0^{r_0} (\frac{w(r)}{r^2}
    \sqrt{1 + f(r)(\partial_r x)^2} - \frac{w(0)}{r^2} - \frac{w'(0)}{r}) \, dr - 2\frac{g}{r_0}w(0) + 2 g w'(0) ln(r_0)\\
    &= 2g \int_0^{r_0} (\frac{w(r)}{r^2}
    \sqrt{1 +  \frac{f(r_0) w^2(r_0)/r_0^4}{f(r)  w^2(r)/r^4 - f(r_0) w^2(r_0)/r_0^4}} - \frac{w(0)}{r^2} - \frac{w'(0)}{r}) \, dr\\
    &- 2\frac{g}{r_0}w(0) + 2 g w'(0) ln(r_0).
  \end{aligned}
\end{equation}
In the Andreev-Zakharov model, $g = 0.176$, which is related to the string tension, $w(r) = e^{s r^2}$, and $s$ is set to 0.45.  At vanishing temperature, we just need to set $f(r) = 1$. In the next sections, we will use two different types of neural network to construct a holographic model from lattice results at vanishing temperature.

\section{Modeling Holographic QCD with MLP}\label{sec3}

\begin{figure}
    \centering
    \includegraphics[width=16cm]{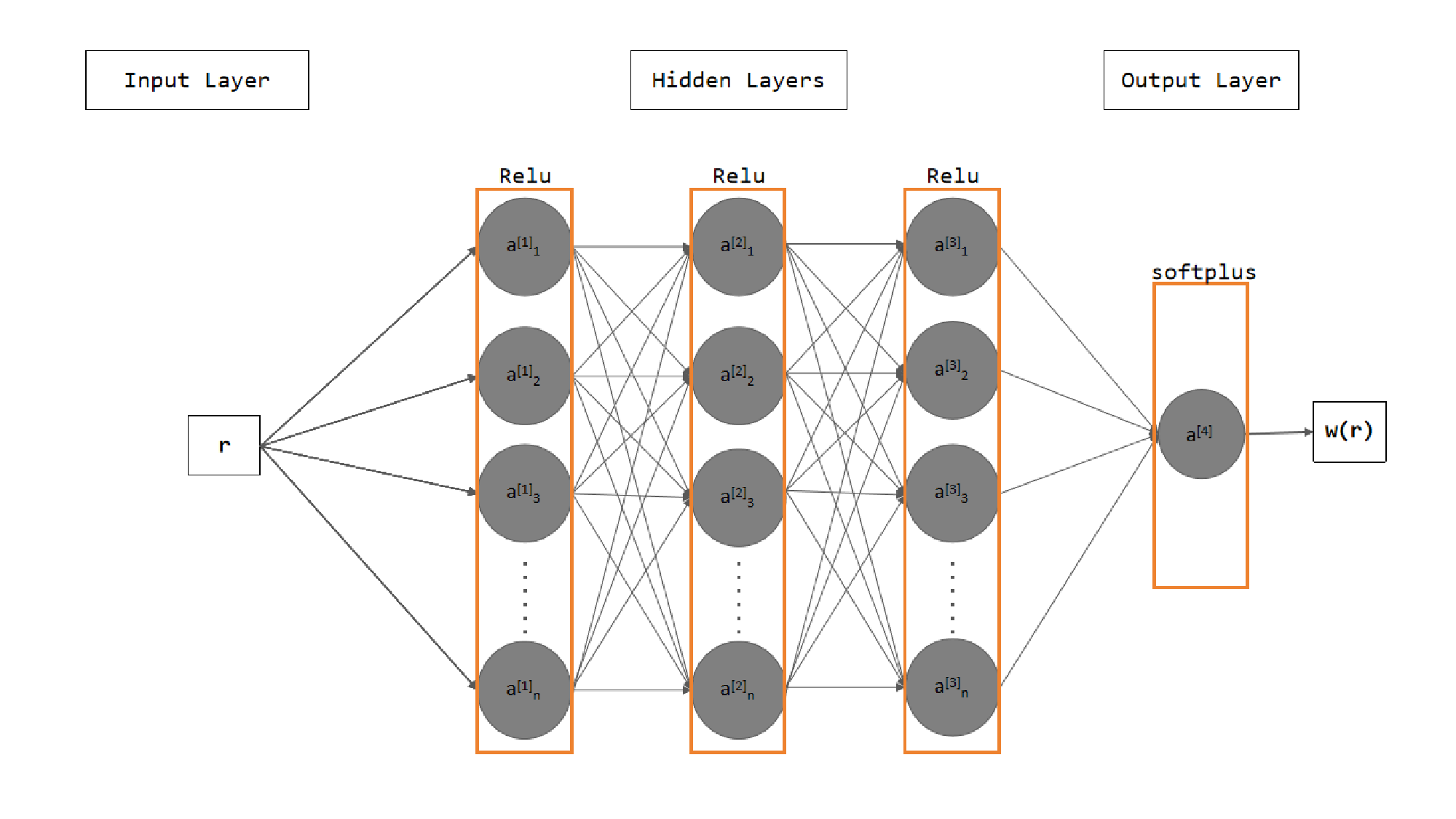}
    \caption{A procedure for reconstructing the \( w(r) \) based on lattice data.}
    \label{obdnn}
\end{figure}

\begin{figure}
  \centering
  \includegraphics[width=16cm]{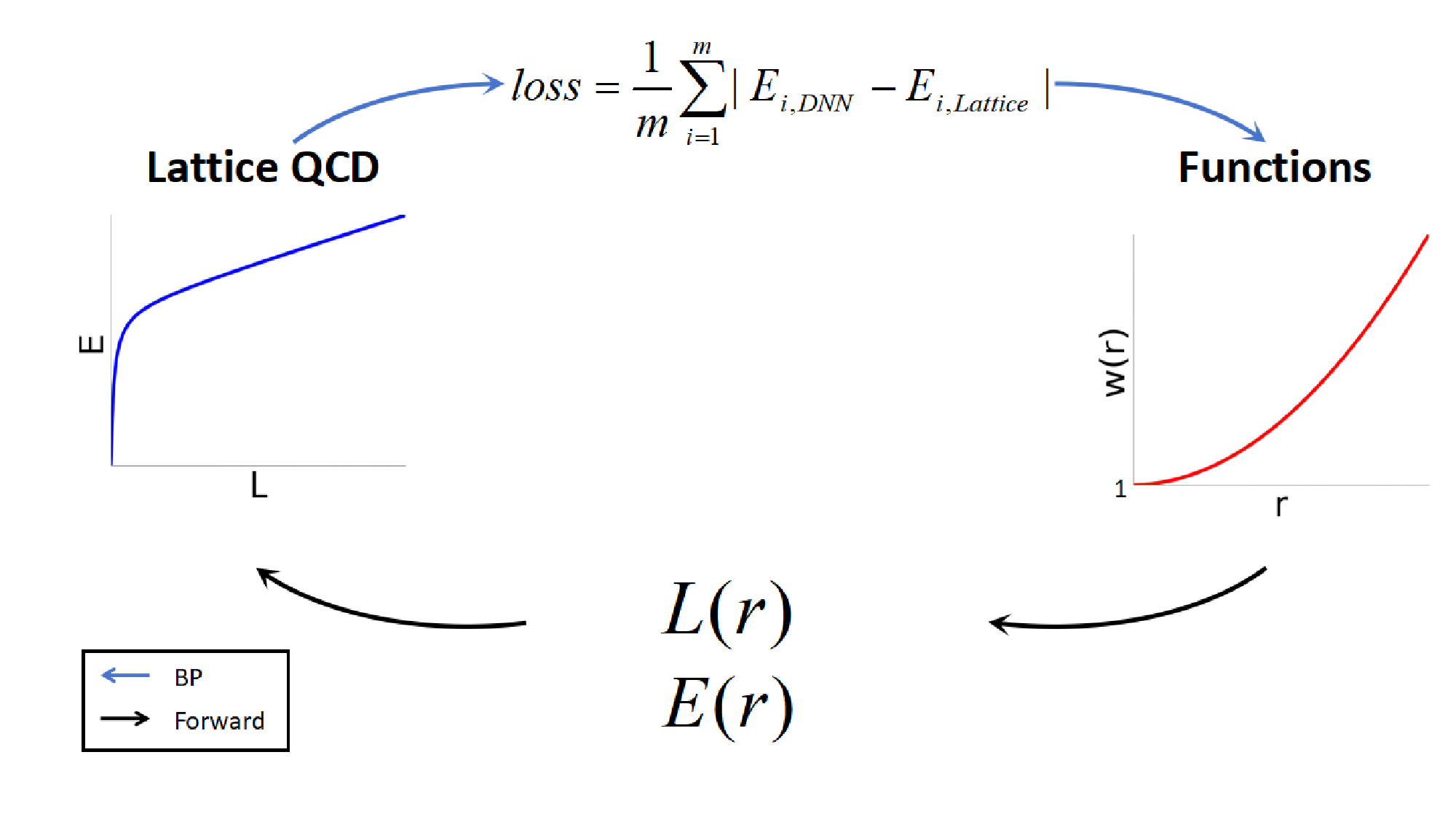}
  \caption{The diagram illustrates MLP architecture designed to model the deformation factor $w(r)$ within the framework of the Andreev-Zakharov model. }
  \label{mlpla}
\end{figure}

The Fig. \ref{obdnn} shows the MLP architecture used in this work for representing the unknown function \(x(r)\). This MLP has three hidden layers with 64, 128 and 64 neurons. The $a^{[1]}_n$, $a^{[2]}_n$ and $a^{[3]}_n$ are the parameters of MLP, which can be optimized using gradient descent algorithm. The end of each hidden layer is the ReLU activation function. In the output layer, we use the softplus activation function. Finally, one can obtain the real output $w(r)$. Once $w(r)$ is given, we can use the 50 points of Gaussian-Legendre to numerically compute the $L(r)$ and $E(r)$ respectively following Eq. (\ref{L}) and Eq. (\ref{E_eq}) as shown in the Fig. \ref{mlpla}. The training objective is to minimize the Mean Absolute Error (MAE) between the $E$ from the MLP-learned $w(r)$ and the lattice QCD result \cite{Cheng:2007jq}. Besides, we also notice that the boundary condition of AdS space $w(r \rightarrow 0 ) \rightarrow 1$, which is introduced into the loss function as the physical constraint term.

\begin{figure}
  \centering
  \includegraphics[width=16cm]{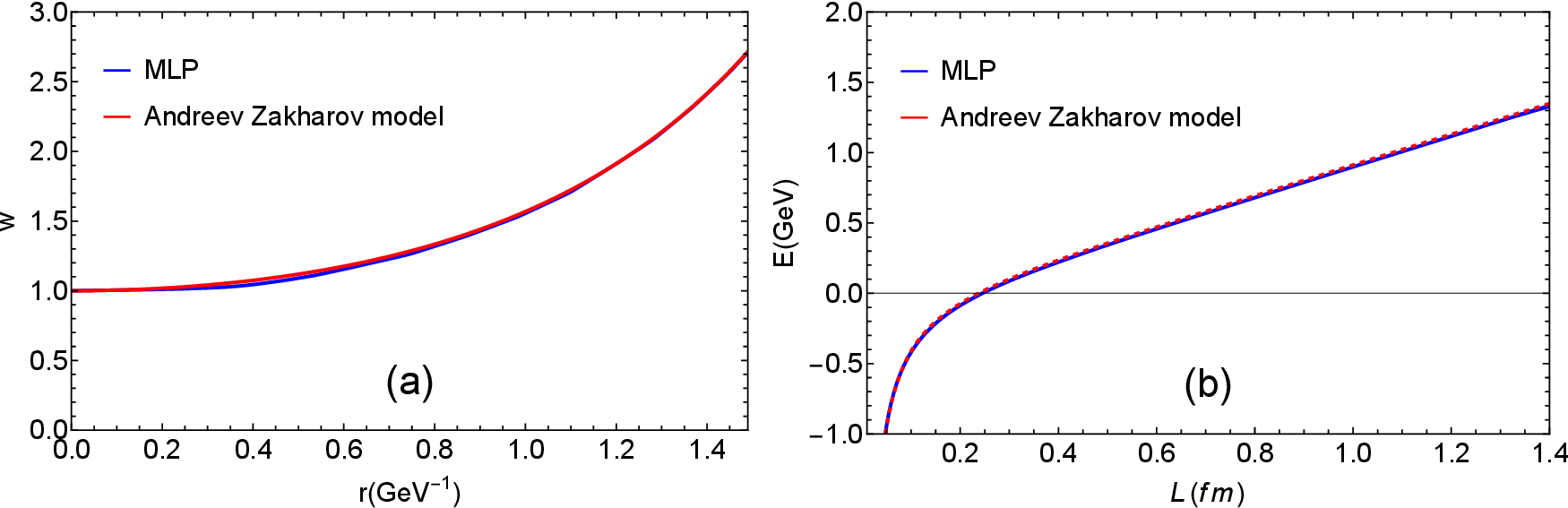}
  \caption{The outcome of the MLP training demonstrates a remarkable alignment with the Andreev-Zakharov model.}
  \label{mlp_andreev}
\end{figure}

Before proceeding, we employ the Andreev-Zakharov model to verify the usage of the above MLP. We input the potential data generated by the Andreev-Zakharov model into the MLP. As illustrated in Fig. \ref{mlp_andreev}, the results demonstrate consistency between $w(r)$ and $E(L)$ from the MLP and the Andreev-Zakharov model, well confirming the validity of our method with MLP.

\begin{figure}
    \centering
    \includegraphics[width=16cm]{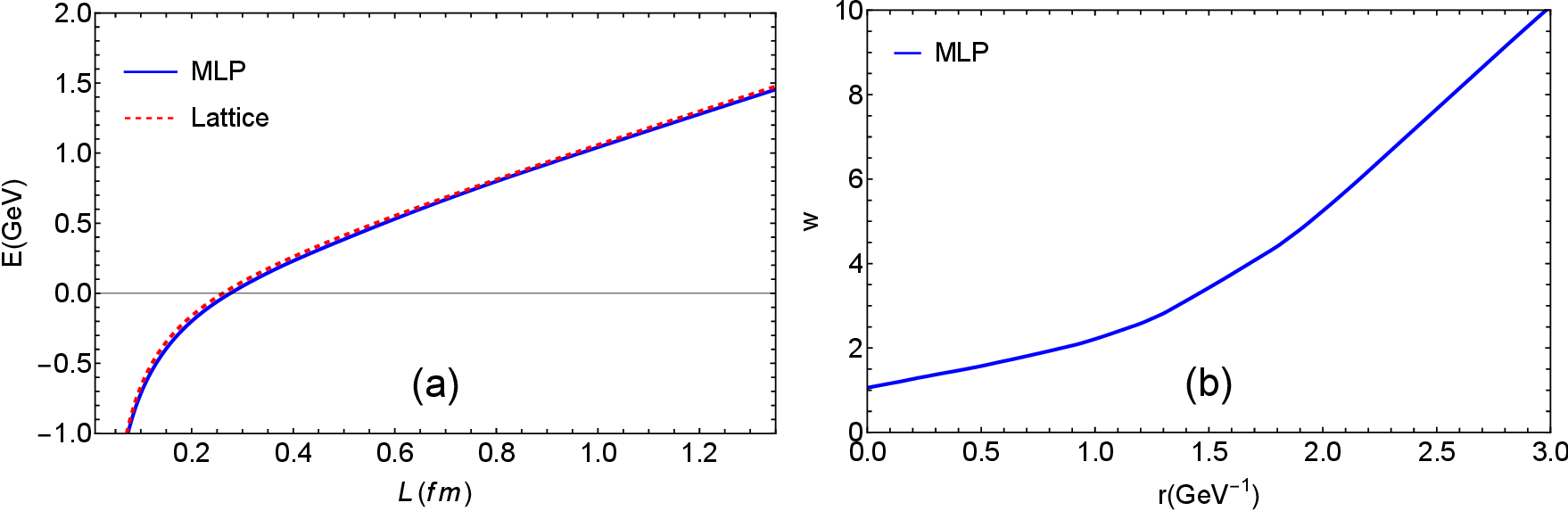}
    \caption{MLP performance on the training dataset illustrating the relationship between $L$ and $E$. Reconstruction of the function $w(r)$ from MLP. $g$ is set to be 0.176.}
    \label{py}
\end{figure}

In Fig. \ref{py} (a), the performance of the neural network is depicted on the training data set, and the curve indicates that the network's ability to represent the function $E(L)$ achieves a high degree of precision, in line with theoretical expectations. After successfully training on the data for $E(L)$, we then reconstructed the function $w(r)$, and the result is shown in Fig. \ref{py} (b). It is evident that the neural network can approximate the exponential trend of $w(r)$ with considerable accuracy and also satisfies the condition $w(r) \rightarrow 1$ at $r = 0$, validating the effectiveness of our model at zero temperature. The heavy quark potential changes sign due to the interplay between the Coulomb term and the linear confinement term in the Cornell potential. At short distances, the negative Coulomb term dominates the potential energy, reflecting the attractive force between the quark and antiquark. This negative potential indicates that the energy of the bound quark-antiquark pair is lower than that of two free quarks, signifying a stable bound state. As the separation distance L increases, the linear term representing the confining potential becomes significant. This term contributes positively to the potential energy and increases linearly with distance, modeling the phenomenon of quark confinement at larger scales. The positive contribution from the linear term eventually outweighs the negative Coulomb term, causing the total potential energy to become positive.

\begin{figure}
    \centering
    \includegraphics[width=16cm]{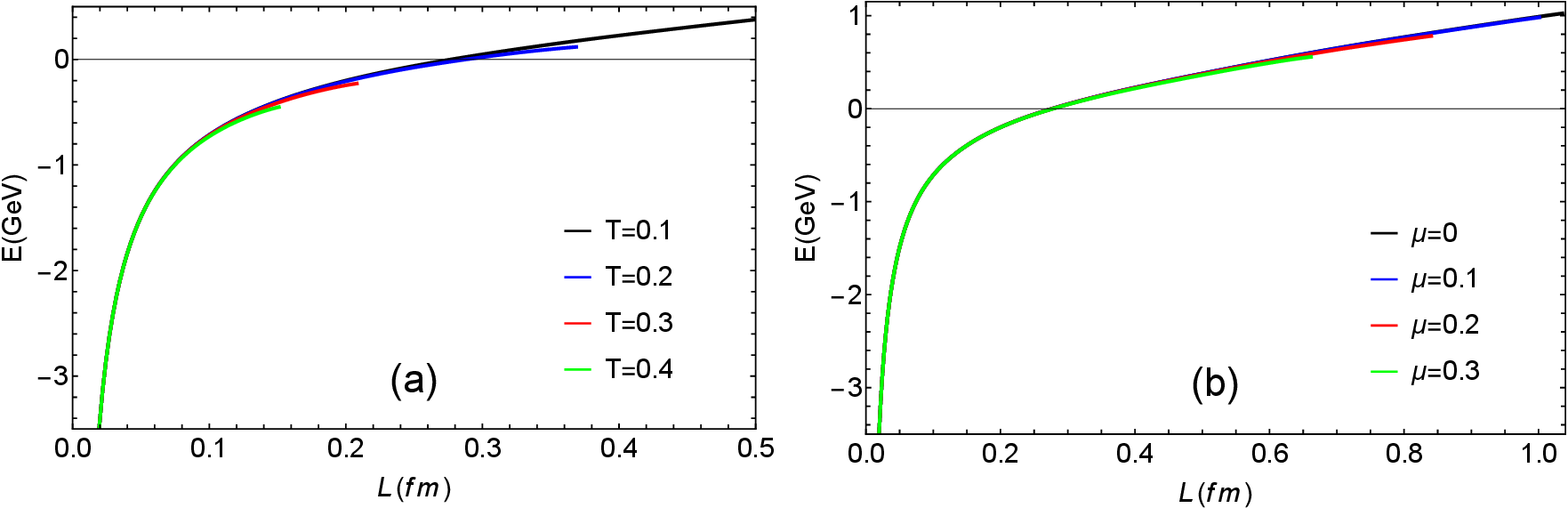}
    \caption{(a) Dependence of the potential energy \(E\) of quark-antiquark pairs on the quark separation distance \(L\) at different temperatures when \( \mu = 0 \). (b) At \( T = 0.1 \), the dependence of the potential energy \(E\) of quark-antiquark pairs on the quark separation distance \(L\) under different chemical potentials. The unit of \(L\) is fm, \( \mu \) is in GeV, \( r \) is in \( \text{GeV}^{-1} \), and \( E \) is in GeV.}
    \label{tannc}
\end{figure}

Next, we used the numerical solution of $w(r)$ from the MLP to calculate the potential energy at finite temperature and the chemical potential. Fig. \ref{tannc} (a) shows that the finite temperature slightly decreases the linear component of the potential and has minimal influence on the Coulombic component. As the temperature increases, the potential vanishes at a smaller distance, indicating that heavy quarks dissociate. This behavior is consistent with the findings in \cite{Bala:2021fkm}, suggesting that the numerical solution for $w(r)$ obtained through the neural network is reliable. Fig. \ref{tannc} (b) also shows a similar trend, demonstrating that as the chemical potential increases, quark pairs begin to screen at a smaller distance, resulting in the quarks becoming free. However, the effect of chemical potential is less pronounced than that of temperature.

\section{Modeling Holographic QCD with KAN}\label{sec4}
In this section, KAN demonstrate significant efficacy. While MLP have fixed activation functions at nodes ('neurons'), KAN feature learnable activation functions on edges ('weights'). Remarkably, KAN eliminate linear weights entirely -- each weight parameter is replaced by a univariate function parameterized as a spline \cite{Liu:2024swq}.

\begin{figure}
  \centering
  \includegraphics[width=15cm]{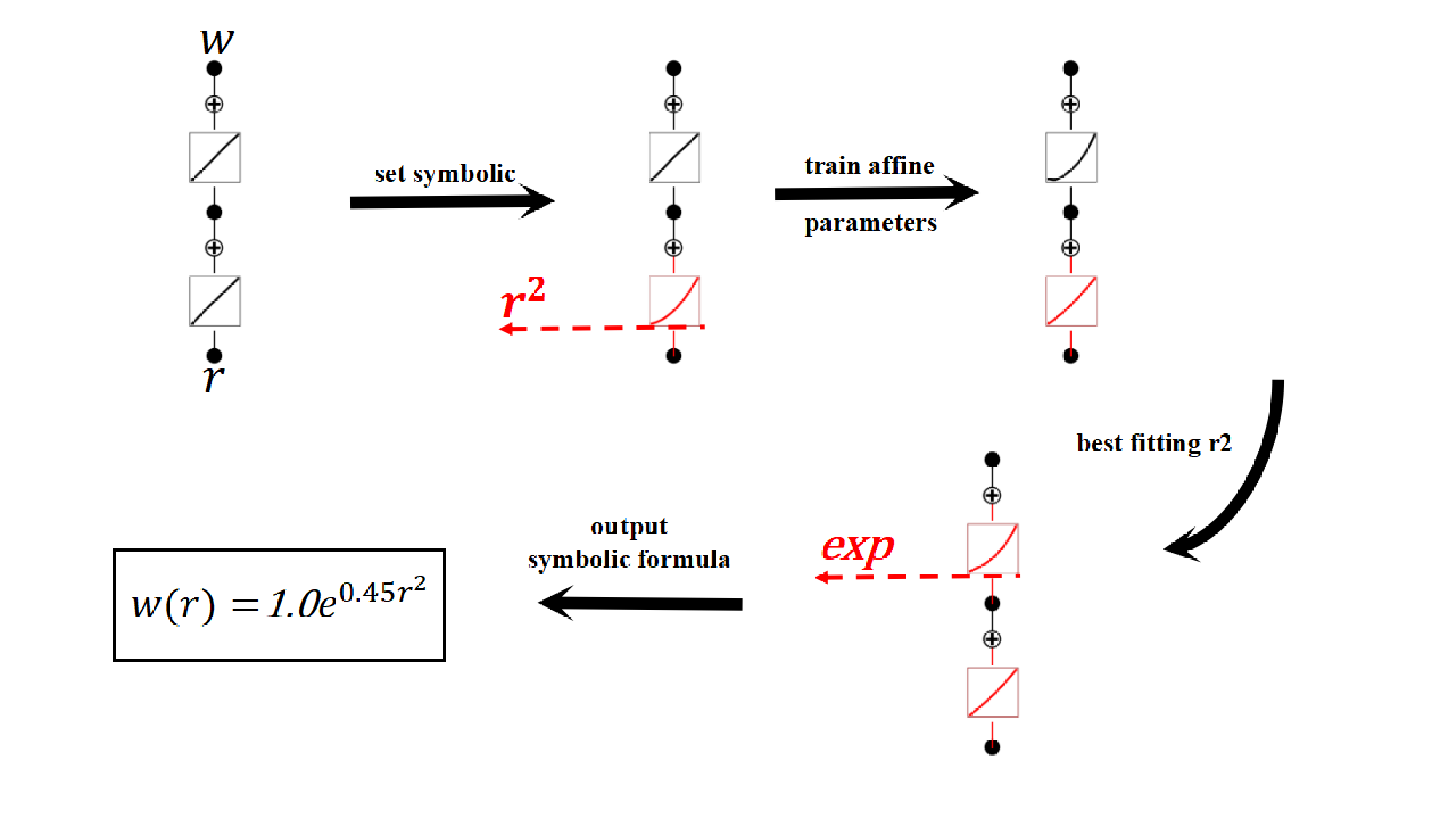}
  \caption{The KAN structure and its application to reproduce the Andreev-Zakharov model.}
  \label{kanandreev}
\end{figure}

\begin{figure}
  \centering
  \includegraphics[width=16cm]{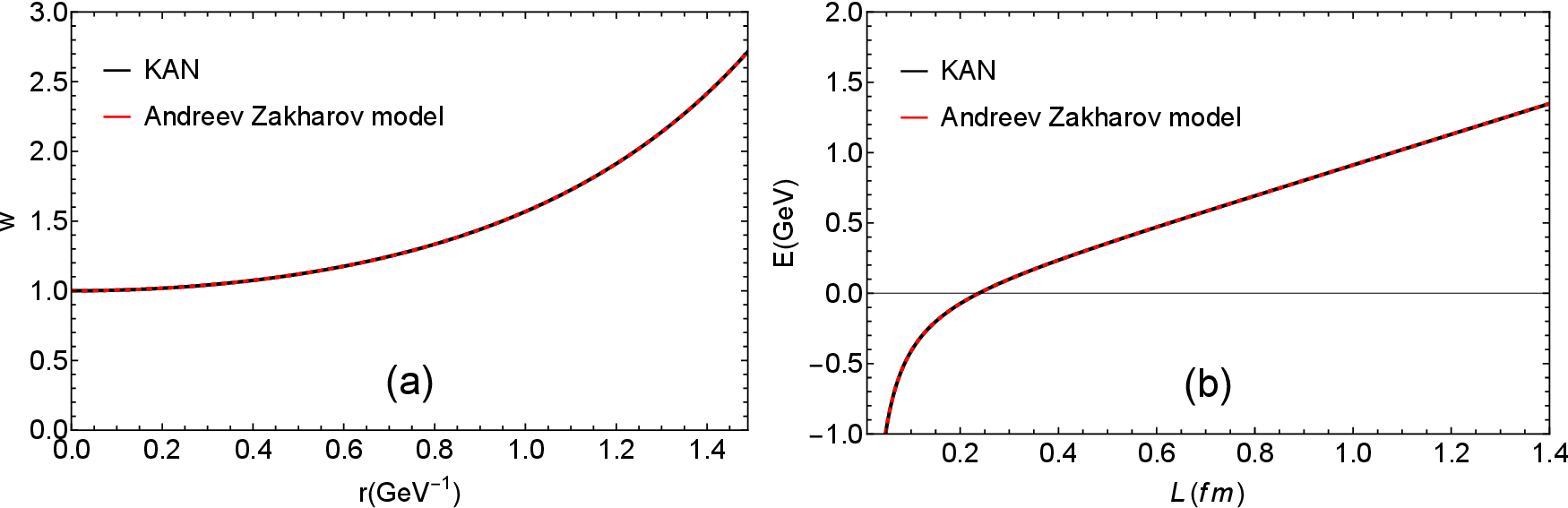}
  \caption{The outcome of the KAN training demonstrates a remarkable alignment with the Andreev-Zakharov model.}
  \label{kanandreev_we}
\end{figure}

\begin{figure}
  \centering
  \includegraphics[width=15cm]{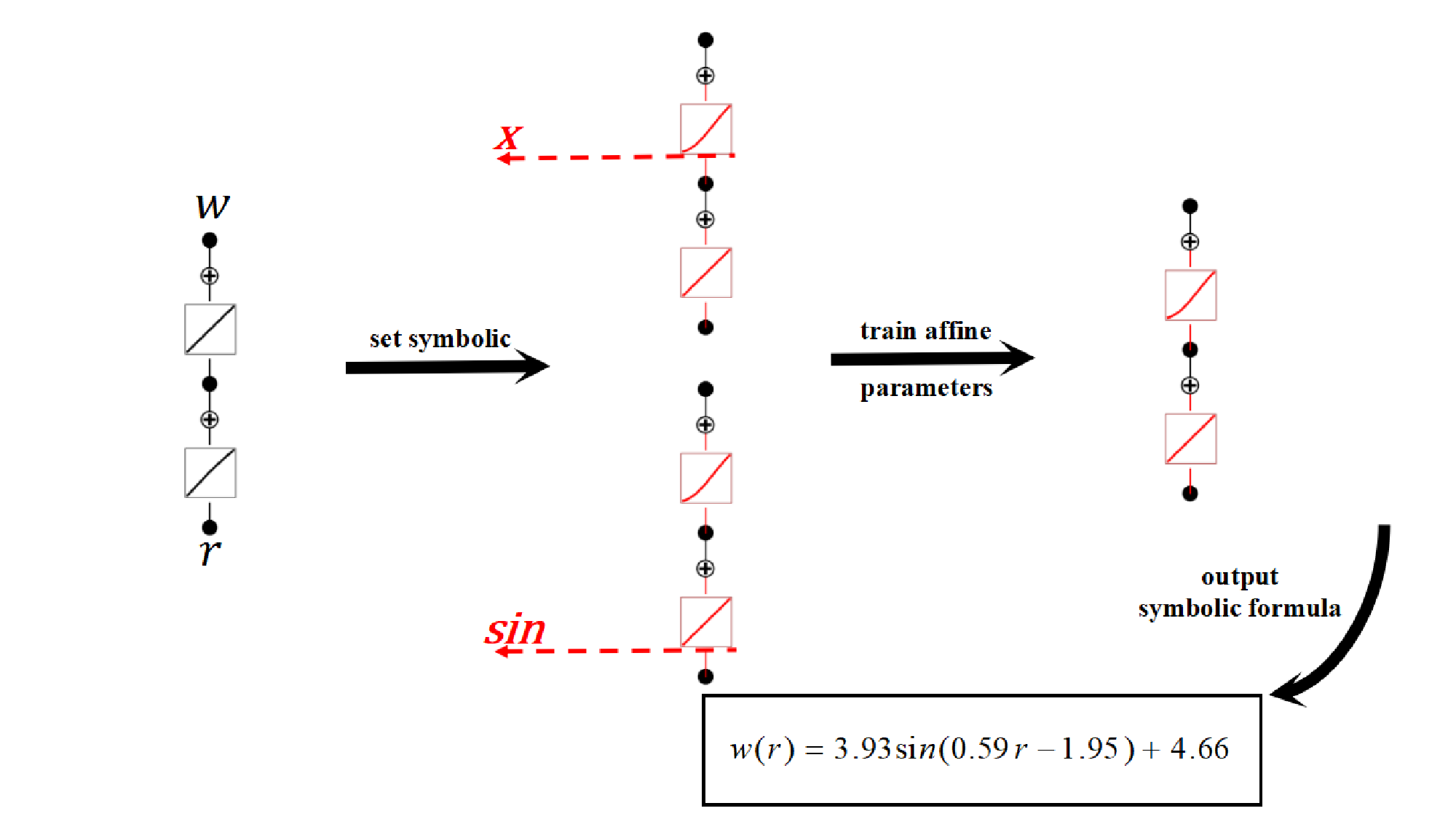}
  \caption{This diagram shows KAN architecture. In the first layer with a \textit{sin} activation function in the spline model; in the second layer with an \textit{x} activation function; $g$ is given as 0.4947 by KAN.}
  \label{kannet}
\end{figure}

First, we will check the validity of KAN. The Andreev-Zakharov model serves as the target model with $w(r) = e^{0.45 r^2}$. Utilizing the expression formula of the KAN, given by
\begin{equation}
  \label{eq:kan}
  w(r) = w(r_1, \cdots, r_n) = \sum_{q=1}^{2n+1} \Phi_q \left( \sum_{p=1}^{n} \varphi_{q,p}(r_p) \right).
\end{equation}
We can ascertain that the neural network node is configured as (1, 3, 1) and $n = 1$. This configuration employs two spline functions to achieve the representation of Andreev-Zakharov model as shown in Fig. \ref{kanandreev}.
By fitting this model with the KAN, we obtain a perfectly fitting result
\begin{equation}
  \label{Andreevmode}
  w(r) = 1.0 e^{0.45  r^2}.
\end{equation}
The result confirms the validity of KAN as shown in Fig. \ref{kanandreev_we}.
Consistent with our previous approach, we hypothesize that $w(r)$ is a specific function derived from lattice data, which we model using a two-layer structure in Fig. \ref{kannet}. Accordingly, we have constructed the KAN with an architecture of (1, 1, 1), indicating the number of nodes in the input, hidden, and output layers, respectively. The function \( w(r) \) trained by the KAN is
\begin{equation}
  \label{eq:kanw}
  w(r) = 3.93 \cdot \sin(0.59 \cdot r - 1.95) + 4.66
\end{equation}

\begin{figure}
  \centering
  \includegraphics[width=12cm]{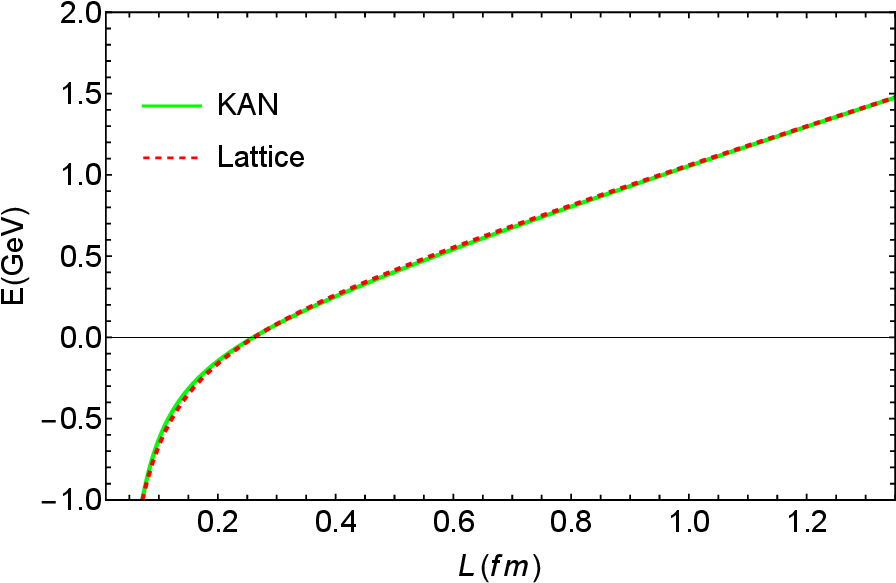}
  \caption{Comparison of potential energy calculations using the KAN model against lattice data.}
  \label{kane}
\end{figure}

\begin{figure}
  \centering
  \includegraphics[width=16cm]{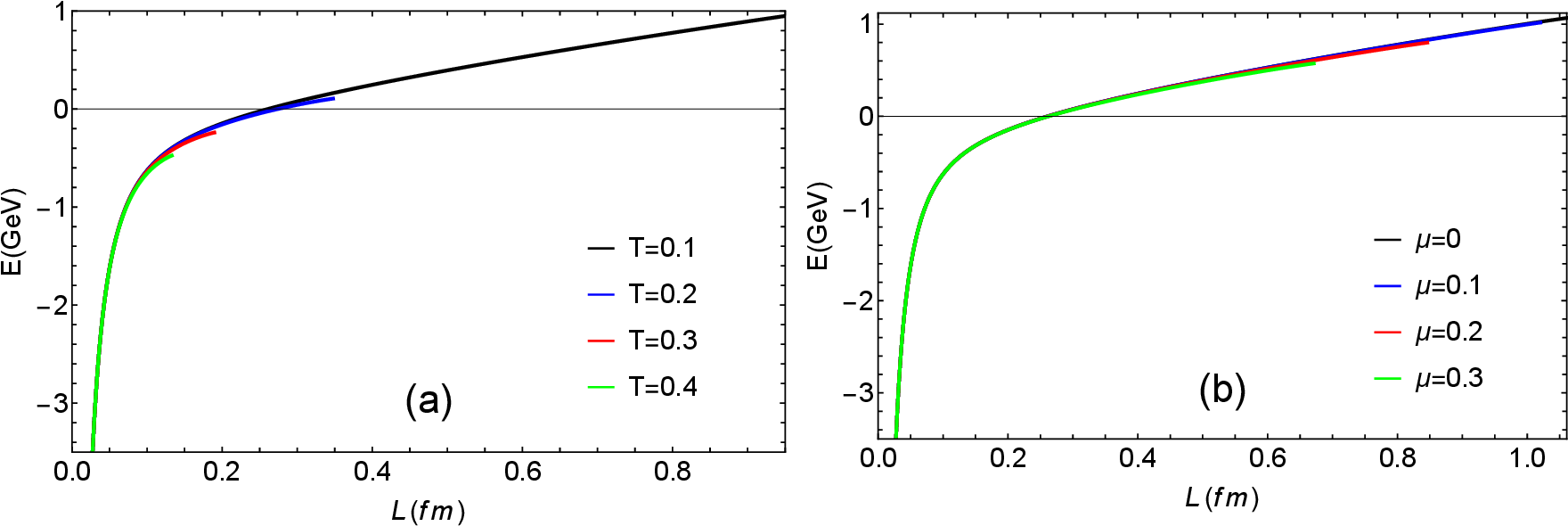}
  \caption{(a) The potential energy \(E\) of quark-antiquark pairs as a function of the quark separation distance \(L\) at various temperatures when the chemical potential \( \mu = 0 \), analyzed within the KAN framework for finite temperature and chemical potential. (b) The variation of potential energy \(E\) as a function of the quark separation distance \(L\) at a fixed temperature \( T = 0.1 \) for different chemical potentials. The units are as follows: \(L\) is measured in fm, \( \mu \) in GeV, \( r \) in \( \text{GeV}^{-1} \), and \( E \) in GeV.}
  \label{kanct}
\end{figure}

According to Eq. (\ref{eq:kanw}), the potential energy plot is shown in Fig. \ref{kane}. It can be observed that the KAN model can also fit the lattice results well while giving an analytical expression.
In Fig. \ref{kanct}, we calculate the relationship between the potential energy \(E\) of quark-antiquark pairs and their separation distance \(L\) under varying temperatures and chemical potentials, within the framework of the KAN. The left side (a) shows how the potential energy depends on the quark separation distance \(L\) at different temperatures with the chemical potential \( \mu = 0\). The right side (b) depicts the variation of potential energy \(E\) as a function of the quark separation distance \(L\) at a fixed temperature \( T = 0.1\) and different chemical potentials. The results demonstrate a qualitative behavior consistent with our previous findings. The most important constraint on \( w(r) \) is \( w(0) \rightarrow 1 \), which is a requirement of asymptotic $\rm AdS_5$ spacetime for the UV regime by holography. At vanishing temperature and chemical potential shown in Fig. \ref{kane}, the distance between the end points goes to infinity, which means the heavy-quark pair is permanently confined. As we can see in Fig. \ref{kanct}, when the distance between the end points of the string is large, the quark-antiquark pair will be screened, meaning the quarks become free. The string will fall inside the horizon at high temperatures and chemical potentials. The maximum value represents the screening distance. Thus, our results are physically plausible.

\section{The Comparison of Models}\label{sec5}
\begin{figure}
    \centering
    \includegraphics[width=16cm]{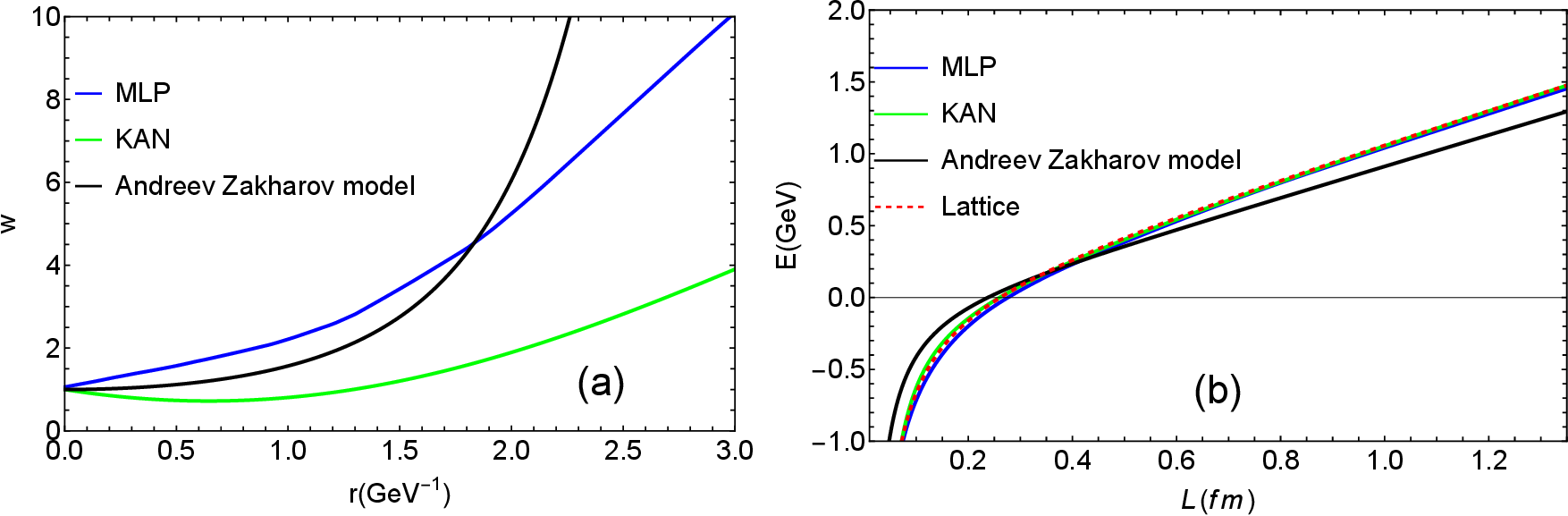}
    \caption{(a) $w(r)$ are represented in MLP, KAN, and Andreev-Zakharov model. (b) The potential energy calculated by MLP, KAN, Andreev-Zakharov model, and lattice QCD.}
    \label{comppp}
\end{figure}

In this section, we make a comparison of all models with lattice QCD data. The results for the deformation factor $w(r)$ are shown in Fig. \ref{comppp} (a), which illustrates that the outcomes of KAN and MLP are in close agreement. Fig. \ref{comppp} (b) demonstrates that both KAN and MLP models fit the lattice data well. In contrast, the Andreev-Zakharov model exhibits a deviation from the lattice QCD data. Our study affirms the validity of the inverse problem approach, which involves using known lattice or experimental data to refine the theoretical model.

\section{Summary}\label{sec6}
We employ MLP and KAN to extract the deformation factor within the holographic model from lattice QCD results. Our paper demonstrates the effectiveness of both MLP and KAN in addressing this inverse problem. In particular, KAN are capable of extracting analytical solutions for the model, as opposed to the numerical solutions provided by MLP. We first utilize the Andreev-Zakharov model as a benchmark for KAN, which shows that KAN can accurately reproduce the model's behavior and analytical expression. Subsequently, we apply KAN to extract the deformation factor directly from lattice QCD data, yielding an analytical solution that fits the lattice QCD results well. Lastly, we examine the heavy-quark potential at finite temperature and the chemical potential, revealing the numerical consistency between the KAN and MLP-based inverse extraction in this problem.

The results of the study provide an effective example of using machine learning methods to solve complex physical problems. They also propose new directions for subsequent research, including how to overcome the limitations of existing methods and how to further enhance the performance and interpretability of models. Our work offers a valuable approach for refining theoretical models using data from lattice QCD or experiments, employing both MLP and KAN.

\section*{Acknowledgments}
This work is supported by the Natural Science Foundation of Hunan Province of China under Grants No. 2022JJ40344, the Research Foundation of Education Bureau of Hunan Province, China under Grant No. 21B0402, Open Fund for Key Laboratories of the Ministry of Education under Grants No. QLPL2024P01, the CUHK-Shenzhen university development fund under grant No. UDF01003041 and UDF03003041, and Shenzhen Peacock fund under No. 2023TC0179.
\section*{References}
\bibliography{ref}

\end{document}